\documentclass[3p,times,procedia]{elsarticle}
\usepackage{nupha_ecrc}

\volume{00}

\firstpage{1}

\journalname{Nuclear Physics A}

\runauth{A. Ohlson}

\jid{nupha}

\jnltitlelogo{Nuclear Physics A}

\usepackage{amsmath}
\usepackage{amssymb}

\newcommand{\ptn}{p_{\rm{T}}}

\newcommand{\pp}{{$p$+$p$}}
\newcommand{\pPb}{{$p$+Pb}}
\newcommand{\PbPb}{{Pb+Pb}}
\newcommand{\dAu}{{$d$+Au}}
\newcommand{\HeAu}{{$^3$He+Au}}
\newcommand{\pAu}{{$p$+Au}}
\newcommand{\pAl}{{$p$+Al}}
\newcommand{\pA}{{$p$+$A$}}
\newcommand{\AuAu}{{Au+Au}}

\newcommand{\sqrts}{\sqrt{s_{\rm{NN}}}}

\begin{document}

\begin{frontmatter}

\dochead{XXVIth International Conference on Ultrarelativistic Nucleus-Nucleus Collisions\\ (Quark Matter 2017)}

\title{Collective behavior in small systems}

\author[label1]{Alice Ohlson}
\address[label1]{Physikalisches Institut, Ruprecht-Karls-Universit\"{a}t Heidelberg, Germany}

\begin{abstract}
The presence of correlations between particles significantly separated in pseudorapidity in proton-proton and proton-nucleus collisions has raised questions about whether collective effects are observed in small collision systems as well as in heavy-ion collisions.  The quantification of these long-range correlations by $v_n$ coefficients is of particular interest.  A selection of the latest $v_n$ measurements is presented, including results from the recent \dAu{} beam energy scan at RHIC where a significant non-zero $v_2$ is measured down to low center-of-mass energies ($\sqrts$ = 39 GeV).  Results from a collision system scan -- comprising \pAu{}, \dAu{}, and \HeAu{} collisions -- are also shown to address the role of the initial nuclear geometry in the final state anisotropy.  Finally, the challenge of measuring multi-particle cumulants, particularly $c_2\{4\}$, in \pp{} collisions is discussed, and new methods for reducing the effects of non-flow are shown to produce a more robust measurement of $v_2\{4\}$ in \pp{} collisions.   
\end{abstract}


\end{frontmatter}



\section{Introduction}

The surprising discovery of the ``ridge'' -- a correlation between particles which are significantly separated in pseudorapidity ($\eta$) -- in \pp{} and \pA{} collisions opened many questions about the interpretation of these correlation structures both in heavy-ion collisions and in smaller systems.   Causality arguments suggest that correlations between particles separated in $\eta$ should originate at early times in the evolution of the collision, either in the initial state or in the initial energy distribution.  It is suggested that collective interactions are needed to transform the early spatial correlations into the observed final-state momentum-space correlations.  In heavy-ion collisions, the presence of long-range ridges has typically been attributed to the collective hydrodynamic behavior of the medium produced in such collisions, but it is unclear whether the ridges in small systems imply hydrodynamics or if there are other physical mechanisms which produce similar structures.  In this proceedings, to avoid making reference to a specific physical explanation for the ridge, collectivity is defined as multiple particles correlated across rapidity due to a common source.  

The ridge in small systems was first observed in an analysis of two-particle angular correlations in high multiplicity \pp{} collisions at $\sqrt{s}$ = 7 TeV~\cite{CMSppridge}, where it appears as a correlation in azimuthal angle ($\varphi$) between particles across large ranges of pseudorapidity.  The nearside ridge was also observed in \pPb{} collisions at $\sqrts$ = 5.02 TeV~\cite{CMSpPbridge}, and after subtraction of jet-like structures a symmetric ``double ridge'' was observed on the awayside~\cite{doubleridge,ATLASdoubleridge}.  The double ridge structure has been quantified by $v_n$ coefficients, analogous to those studied in heavy-ion collisions.  Since then, the ridge and $v_n$ have been studied in other collision systems and differentially with respect to momentum, particle species, pseudorapidity, etc.  A selection of the latest results is discussed in these proceedings.  

\section{Observables and analysis methods}

There are several analysis techniques currently employed for studying ridges in small (and large) collision systems.  The first is the direct analysis of two-particle correlation functions.  The correlation function, $C(\Delta\varphi,\Delta\eta)$, is defined as the distribution in relative azimuthal angle ($\Delta\varphi = \varphi_{assoc}-\varphi_{trig}$) and relative pseudorapidity ($\Delta\eta = \eta_{assoc}-\eta_{trig}$) for particle pairs consisting of a trigger and associated particle.  In order to study various physical effects, correlation functions can be constructed differentially with respect to properties of the trigger and associated particles (such as their transverse momenta $\ptn$, species, and pseudorapidity) and event properties (such as the centrality or multiplicity of the collision).  Correlation functions in \pp{} collisions are dominated by features characteristic of (mini-)jet production: a nearside peak localized around $(\Delta\varphi,\Delta\eta) = (0,0)$, representing pairs of particles where the trigger and associated particles are fragments of the same jet, and the awayside peak localized around $\Delta\varphi = \pi$ but extended in $\Delta\eta$, representing pairs in which the trigger and associated particles are in back-to-back jets.   In heavy-ion collisions, the same jet structures are observed, on top of additional correlations in $\Delta\varphi$ which are extended in $\Delta\eta$.  These long-range structures in the two-particle correlation function are typically quantified by the coefficients, $v_n$, of a Fourier cosine series in $\Delta\varphi$, 
\begin{equation}
\frac{dN}{d\Delta\varphi} \propto 1+2v_1^{trig}v_1^{assoc}\cos(\Delta\varphi)+2v_2^{trig}v_2^{assoc}\cos(2\Delta\varphi)+2v_3^{trig}v_3^{assoc}\cos(3\Delta\varphi)+...
\label{eq:vn}
\end{equation}
The magnitudes of the $v_n$ coefficients can be extracted directly by fitting the long-range part of $C(\Delta\varphi,\Delta\eta)$ with Eq.~\ref{eq:vn} (truncated at some order $n$).  Measuring $v_2$ and the other $v_n$ components has been critical to determining the properties and dynamics of the medium created in heavy-ion collisions, and likewise the measurement of $v_n$ in small systems is of particular interest.  

One of the challenges in characterizing the bulk dynamics with $v_n$ coefficients is reducing the influence of non-flow effects, such as jet production or resonance decays.  This must be handled with particular care in small collision systems where multiplicities are significantly lower than in heavy-ion collisions and a significant proportion of the particle production is from (mini-)jet fragmentation.  It should be emphasized here that while correlations between particles which do not have a collective origin are commonly called ``non-flow,''  a term which has been adopted from large collision systems into small ones, the use of the word ``non-flow'' does not imply the existence of flow in small systems.  

While fitting the correlation function at large $|\Delta\eta|$ makes it possible to avoid the nearside jet peak, the awayside jet peak can still contribute to the measured $v_n$ coefficients (predominantly to $v_1$).  For this reason, several methods have been utilized to further reduce the effects of non-flow on the measured $v_n$ coefficients.  

One technique for removing correlations due to jet and minijet fragmentation is to subtract the correlations in low-multiplicity events from the high-multiplicity correlation functions.  This procedure relies on the assumption that jet structures (both on the nearside and awayside) are independent of multiplicity, which was shown to be a reasonable approximation in \pPb{} collisions~\cite{minijet}.  The ``template fit'' procedure~\cite{ATLAStemplate} is similar but allows the yields of the jet peaks to vary with multiplicity.  In this method, the correlation function is fit with the following function, 
\begin{equation}
Y^{\textrm{templ}}(\Delta\varphi) = F Y^{\textrm{periph}}(\Delta\varphi)+G(1+2v_{2,2}\cos{2\Delta\varphi)},
\end{equation}
where $Y^{\textrm{periph}}(\Delta\varphi)$ is the correlation function in a selected low-multiplicity bin and $F$, $G$, and $v_{2,2}$ are fit parameters.  
Essentially, the template fit method assumes that the shape of the hard (jet-like) component is independent of multiplicity but allows the yield to vary (although note that the yields of the nearside and awayside peaks must scale together by the same factor $F$).  Residual shapes in the correlation function are fit by a cosine series up to order $n = 2$.  


It is necessary to determine whether the observed correlation structures arise from interactions between only a few particles, or whether they are a true collective effect involving a global correlation among many particles within a single event.  To assess this, the multi-particle cumulants, $c_n\{k\}$, are measured, in which correlations between fewer particles are explicitly removed~\cite{cumulants}.  For example, the two-, four-, and six-particle cumulants are:
\begin{equation}
\begin{split}
c_n\{2\} = \langle\langle 2\rangle\rangle &= \langle\langle \cos{n\left(\varphi_1-\varphi_2\right)}\rangle\rangle,\\
c_n\{4\} = \langle\langle 4\rangle\rangle - 2\langle\langle 2\rangle\rangle^2 &= \langle\langle \cos{n\left(\varphi_1+\varphi_2-\varphi_3-\varphi_4\right)}\rangle\rangle - 2\langle\langle 2\rangle\rangle^2,\\
c_n\{6\} = \langle\langle 6\rangle\rangle - 9\langle\langle 4\rangle\rangle\langle\langle 2\rangle\rangle + 12\langle\langle 2\rangle\rangle^3 &= \langle\langle \cos{n\left(\varphi_1+\varphi_2+\varphi_3-\varphi_4-\varphi_5-\varphi_6\right)}\rangle\rangle - 9\langle\langle 4\rangle\rangle\langle\langle 2\rangle\rangle + 12\langle\langle 2\rangle\rangle^3,\\
\end{split}
\label{eq:cumulants}
\end{equation}
where the double angular brackets $\langle\langle\rangle\rangle$ indicate averages taken over all particles in an event and over all events.  The multi-particle cumulants can be related to the $v_n$ coefficients in the following way: 
\begin{equation}
\begin{split}
v_n\{2\}^2 &= c_n\{2\},\\
v_n\{4\}^4 &= -c_n\{4\},\\
v_n\{6\}^6 &= c_n\{6\}/4.
\end{split}
\label{eq:vncumulants}
\end{equation}
The higher-order cumulants suppress non-flow effects, which typically produce correlations between fewer particles.  As can be seen from Eqs.~\ref{eq:cumulants} and~\ref{eq:vncumulants}, in order to obtain real-valued $v_2\{4\}$ it is necessary for $c_2\{4\}$ to be negative.  Positive values of $c_2\{4\}$ are typically viewed as an indication of the dominance of non-flow over collective effects, and are often measured in regimes where non-flow is expected to be significant such as in low-multiplicity collisions.  

\section{Collision systems and energies}
In this proceedings, data from the Large Hadron Collider (LHC) at CERN and the Relativistic Heavy Ion Collider (RHIC) at Brookhaven National Laboratory are presented.  The proton-proton collision data shown here have been recorded at RHIC at $\sqrt{s}$ = 200 GeV and at the LHC at 7 and 13 TeV.  Asymmetric collisions are used to probe the interactions of individual nucleons with the nuclear parton distribution function as well as how geometrical effects in the initial state propagate to final-state anisotropies.  For these purposes, \pPb{} collisions have been recorded at the LHC at center-of-mass energies of 5.02 and 8.16 TeV, and \dAu{} collisions at $\sqrts$ = 200 GeV were studied at RHIC.  To explore the evolution of the \dAu{} collision system with the center-of-mass energy, a \dAu{} beam energy scan was performed at RHIC where data was collected at $\sqrts$ = 19.6, 39, and 62.4 GeV.  To investigate geometrical effects in small collision systems, \pAu{} collisions are compared with the more elliptical \dAu{} and triangular \HeAu{} systems, all at $\sqrts$ = 200 GeV at RHIC.  

\begin{figure}[t]
\centering
\includegraphics[width=0.44\linewidth]{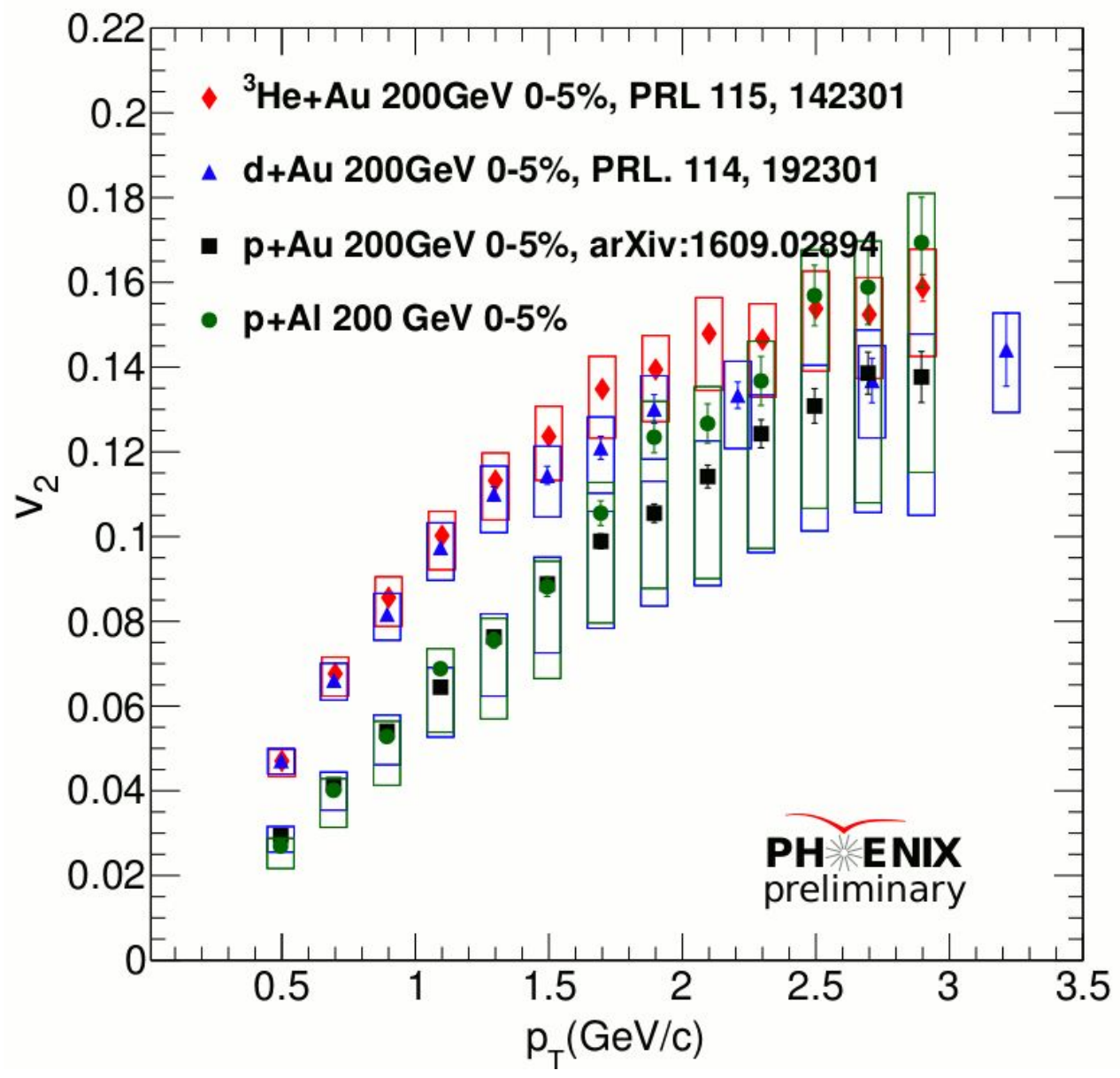}~
\includegraphics[width=0.45\linewidth]{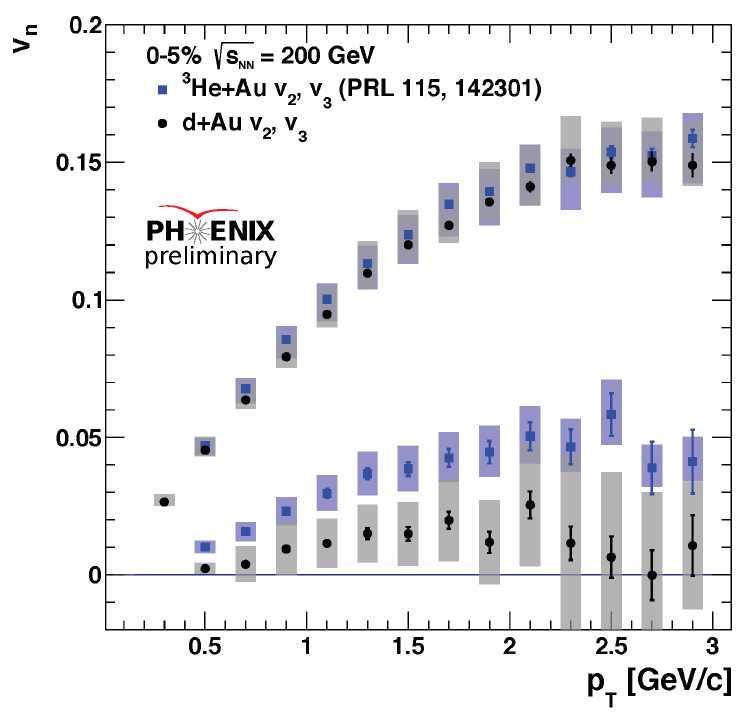}
\caption{The role of initial geometry in the final state azimuthal anisotropy is probed by measuring (left) $v_2$ in \pAl{}, \pAu{}, \dAu{}, and \HeAu{} collisions and (right) $v_3$ in \dAu{} and \HeAu{} at $\sqrts$ = 200 GeV in PHENIX~\cite{PHENIXv2v3pA}.\label{fig:v2v3pA}}
\end{figure}

\section{$v_2$ of light and heavy hadrons in \pPb{}}
The azimuthal anisotropy parameter $v_2$ was first measured in \pPb{} collisions for light hadrons: charged pions, charged and neutral kaons, protons, and lambda baryons~\cite{ALICEPIDridge,CMSPIDridge}.  It was observed that the $v_2$ in \pPb{} shows a similar mass ordering -- with the $\ptn$-differential $v_2$ following one trend for mesons and another for baryons -- as was seen in \PbPb{} collisions.  This species-dependence initially presented a challenge for models which are not based on hydrodynamics, but recent theoretical advances have demonstrated mass ordering in non-hydrodynamic models as well (see~\cite{BjoernTheory} for more detailed discussion and a theory review).  

The $v_2$ of muons from decays of heavy flavor hadrons was measured in \pPb{} collisions by ATLAS using the template fit method~\cite{ATLAShfmuv2}.  The results show that $v_2^{\mu}$ is significantly non-zero, demonstrating that heavy flavor hadrons are also affected by the azimuthal anisotropy of the system.  The observed $v_2^{\mu}$ is smaller than the $v_2$ of unidentified hadrons,  independent of multiplicity, and decreasing with $\ptn$.  Although the decay kinematics and the charm/bottom fraction must be taken into account when quantitatively interpreting these results, future comparisons with theoretical models and with ALICE's $\mu$-h data~\cite{ALICEmuh} will give more information on how light and heavy quarks participate in the azimuthally-asymmetric dynamics of the system.  

\section{$v_n$ response to initial geometry}

In order to probe the connection between initial state geometry and final state azimuthal anisotropy, $v_n$ coefficients were measured in asymmetric systems with different projectiles: protons in \pAu{} and \pAl{} collisions, deuterons in \dAu{} collisions to produce a more elliptical initial energy density distribution, and helium nuclei in \HeAu{} collisions to obtain a more triangular initial geometry~\cite{PHENIXv2v3pA}.  As can be seen in Fig.~\ref{fig:v2v3pA}, the $v_2$ observed in \dAu{} and \HeAu{} is larger than that measured in systems with a colliding proton.  Furthermore, while the $v_2$ values in \dAu{} and \HeAu{} are consistent because of the similar ellipticities ($\varepsilon_2$) of the two systems, $v_3$ is larger in \HeAu{} due to the triangularity ($\varepsilon_3$) of the latter system.  The good correlation between ellipticity (triangularity) and $v_2$ ($v_3$) observed in this system scan demonstrates that the initial geometry of $p$/$d$/$^3$He+Au collisions is well understood.  

\begin{figure}[b]
\centering
\includegraphics[width=0.8\linewidth]{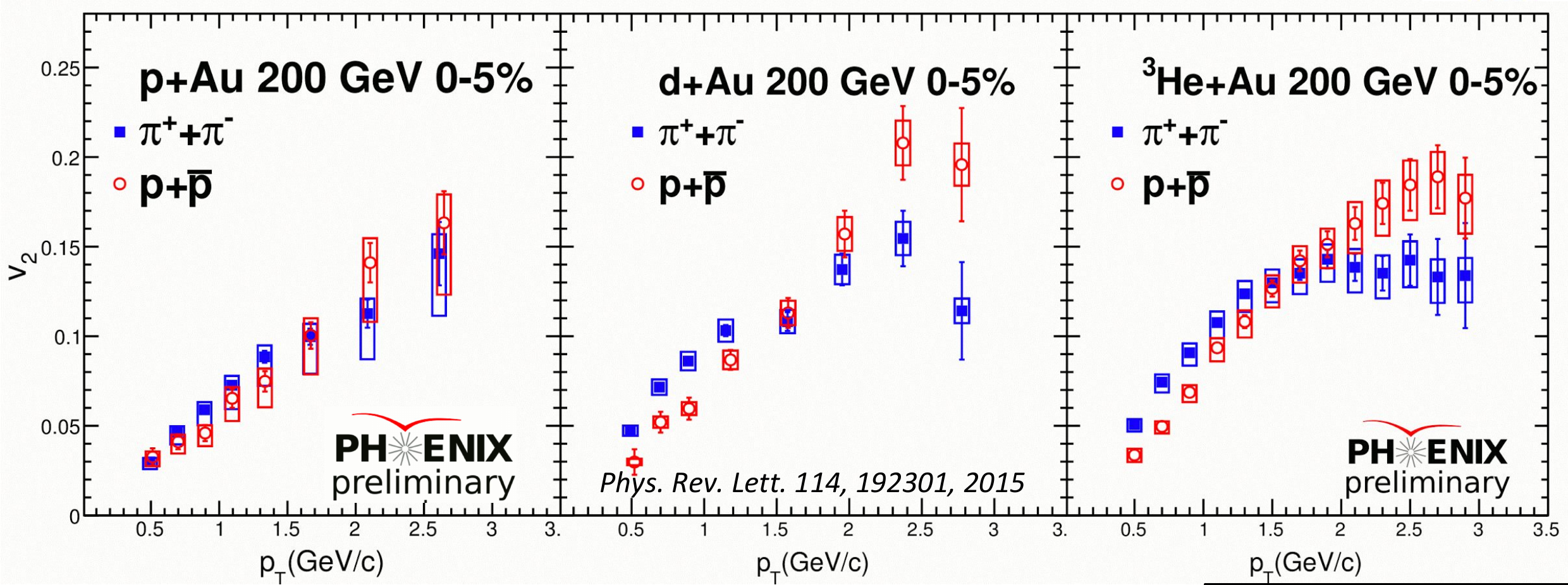}
\caption{The separation between $\ptn$-differential $v_2$ for pions and protons is observed in \pAu{}, \dAu{}, and \HeAu{} collisions at $\sqrts$ = 200 GeV in PHENIX, and is reminiscent of similar mass ordering observed in heavy-ion collisions~\cite{PHENIXv2v3pA}.\label{fig:v2pAPID}}
\end{figure}

Furthermore, the mass ordering in $v_2$ as a function of $\ptn$ observed in \PbPb{} and \pPb{} collisions is seen as well in \pAu{}, \dAu{}, and \HeAu{} collisions, as shown in Fig.~\ref{fig:v2pAPID}.  This observation is consistent with a hydrodynamic picture of \pA{} collisions, but can also be explained by non-hydro models~\cite{BjoernTheory}.  

\section{$v_2$ versus center-of-mass energy}

\begin{figure}[tb]
\centering
\includegraphics[trim={0 0 0 7mm},clip]{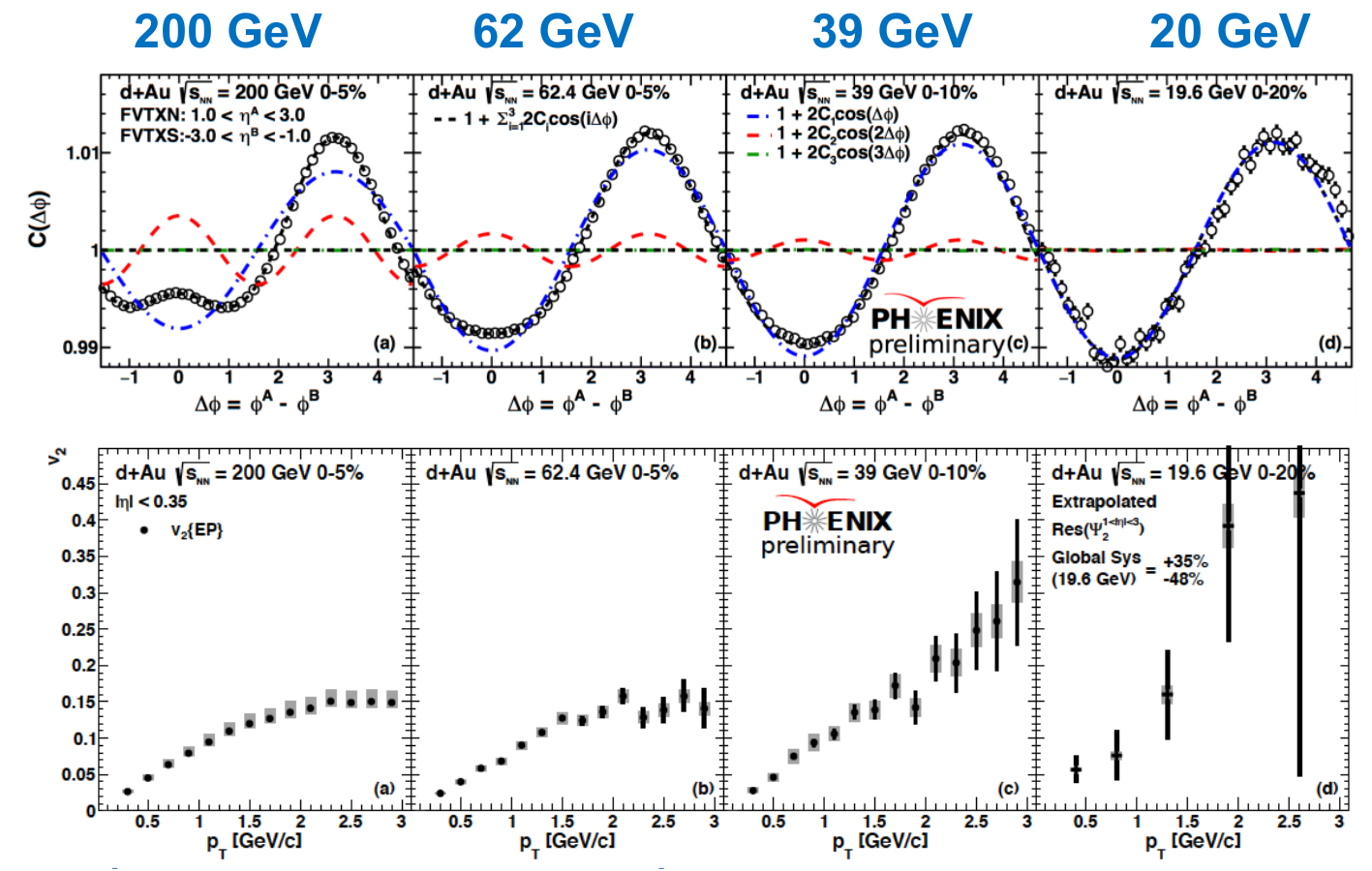}
\caption{Correlation functions $C(\Delta\varphi)$ and $v_2(\ptn)$ from (left) $\sqrts$ = 200 GeV to (right) 19.6 GeV measured in the \dAu{} beam energy scan in PHENIX~\cite{PHENIXv2dAuBES}.\label{fig:v2dAuBES}}
\end{figure}

\begin{figure}[t]
\centering
\includegraphics[trim={0 0 0 5mm},clip]{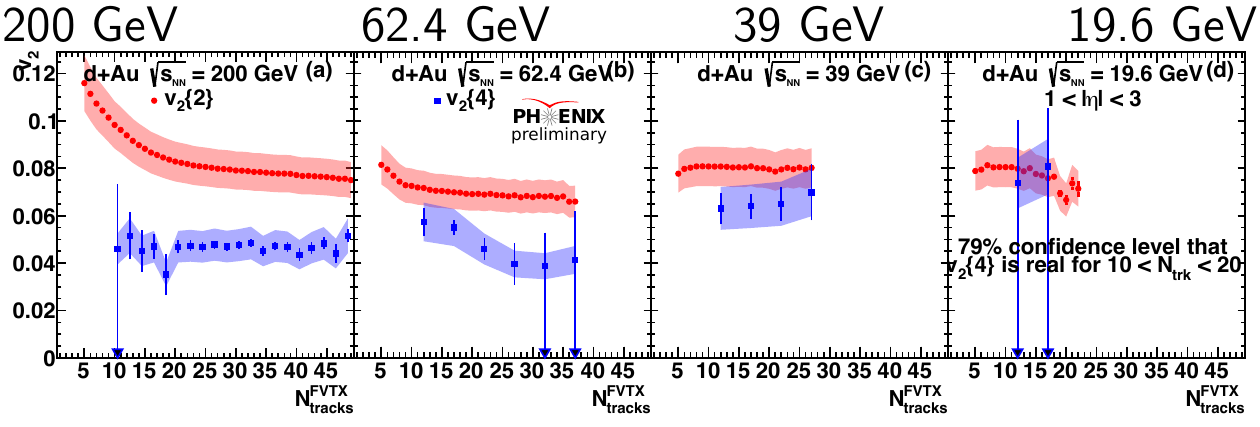}
\caption{The two- and four-particle cumulants, $v_2\{2\}$ and $v_2\{4\}$, were measured as a function of multiplicity in the \dAu{} beam energy scan in PHENIX~\cite{PHENIXv24dAuBES}.  The value of $c_2\{4\}$ is negative, resulting in a measurable $v_2\{4\}$, down to $\sqrts$ = 39 GeV.  \label{fig:v24dAuBES}}
\end{figure}

The correlation functions $C(\Delta\varphi)$ were measured in \dAu{} collisions at $\sqrts =$ 200, 62.4, 39, and 19.6 GeV~\cite{PHENIXv2dAuBES}.  By fitting the long-range correlations ($2 < |\Delta\eta| < 6$), $v_2(\ptn)$ was extracted, as shown in Fig.~\ref{fig:v2dAuBES}.  The data show clear and significant values of $v_2$ down to $\sqrts$ = 39 GeV (in the 19.6 GeV data the measurement is limited by statistical uncertainties).  No significant dependence on the center-of-mass energy is observed; $v_2(\ptn)$ has similar magnitudes at all beam energies recorded in the \dAu{} beam energy scan.  The two- and four-particle cumulant techniques were also used to measure $v_2\{2\}$ and $v_2\{4\}$, respectively, as demonstrated in Fig.~\ref{fig:v24dAuBES}~\cite{PHENIXv24dAuBES}.  The observed $v_2\{4\}$ is significant down to $\sqrts =$ 39 GeV, with a 79\% confidence level that it is real-valued (i.e. $c_n\{4\} < 0$) at 19.6 GeV.  Long-range correlations in $\Delta\eta$ and non-zero $v_n$ coefficients, typically viewed as signatures of collective behavior in heavy-ion collisions, were initially surprising observations in high-multiplicity \pPb{} collisions but have now been measured across a range of \pA{} systems, center-of-mass energies, and multiplicities.  

\section{$v_n$ in \pp{} collisions and the effects of non-flow}

While a variety of techniques have been developed to reduce the influence of non-flow in $v_n$ measurements in \pA{} collisions, the suppression of non-flow in \pp{} collisions is even more challenging.  In these collisions, where event multiplicities are low and particle production is dominated by (mini-)jet fragmentation, even four-particle cumulants are affected by non-flow (and the higher-order cumulants are generally inaccessible in low-multiplicity environments).  Furthermore, it was recently shown that in the presence of significant non-flow and non-Gaussian flow fluctuations $c_n\{4\}$ could have either a positive or negative value, and therefore a negative $c_n\{4\}$ is not necessarily a sign of collectivity~\cite{jiangyong}.  An important test of the (in)sensitivity of $c_2\{4\}$ to non-flow effects is performed by using different multiplicity estimators (such as changing the momentum or pseudorapidity range of the particles used to determine the event multiplicity).  Since each estimator is influenced differently by flow fluctuations and non-flow, the dependence of $c_2\{4\}$ on the multiplicity estimator is indicative of its dependence on non-flow.  Active discussions are ongoing to resolve the discrepancy between the CMS and ATLAS measurements of $c_n\{4\}$ in \pp{} collisions by assessing this dependence on the multiplicity estimator (see for example~\cite{zchen,CMSppcollectivity,ATLASv24subevents}).  

Several new methods were proposed to suppress non-flow effects in measurements of $c_2\{4\}$, including introducing a $|\Delta\eta|$ gap between the pairs of particles in each quadruplet which contributes to the measurement~\cite{ALICEv2pp}, and selecting particles from two or three subevents which imposes that the requirement that the particles entering the measurement are not clustered in pseudorapidity~\cite{jiangyong,ATLASv24subevents}.  (Note that the two-subevent method proposed by ATLAS is equivalent to the measurement of $c_2\{4,|\Delta\eta|>0\}$ used by ALICE.)

\begin{figure}[b]
\centering
\includegraphics[width=0.49\linewidth,trim={11.9cm 0 0 0},clip]{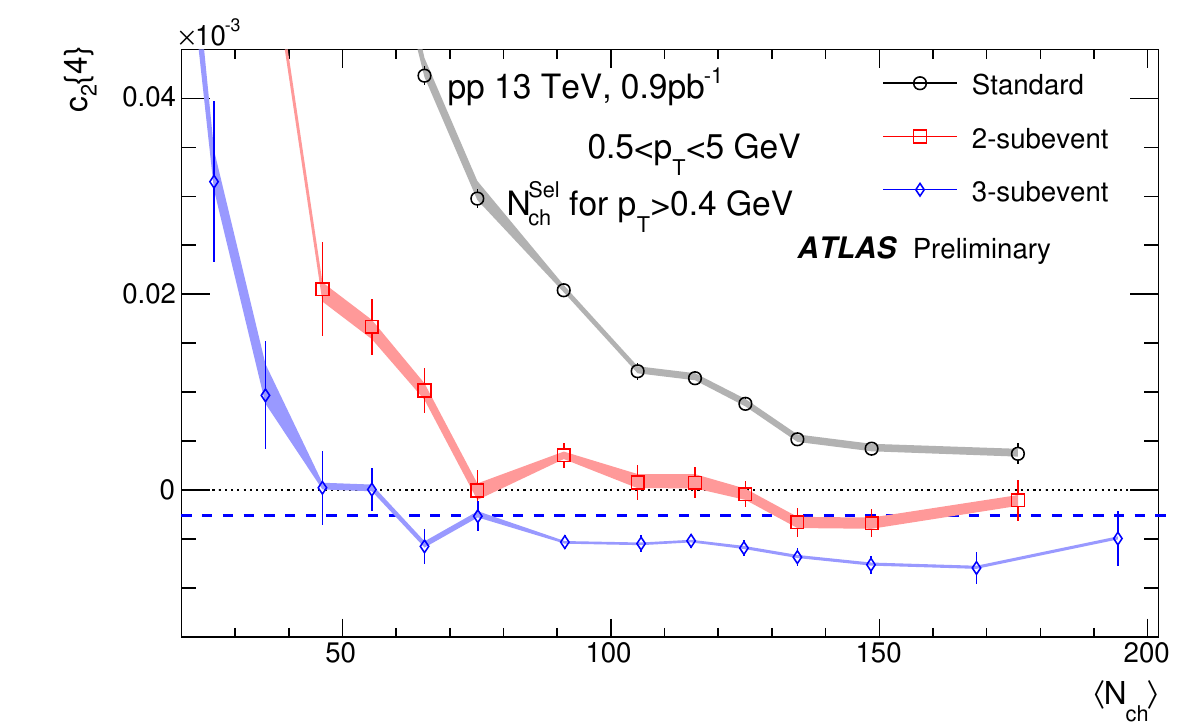}
\includegraphics[width=0.49\linewidth,trim={11.9cm 0 0 0},clip]{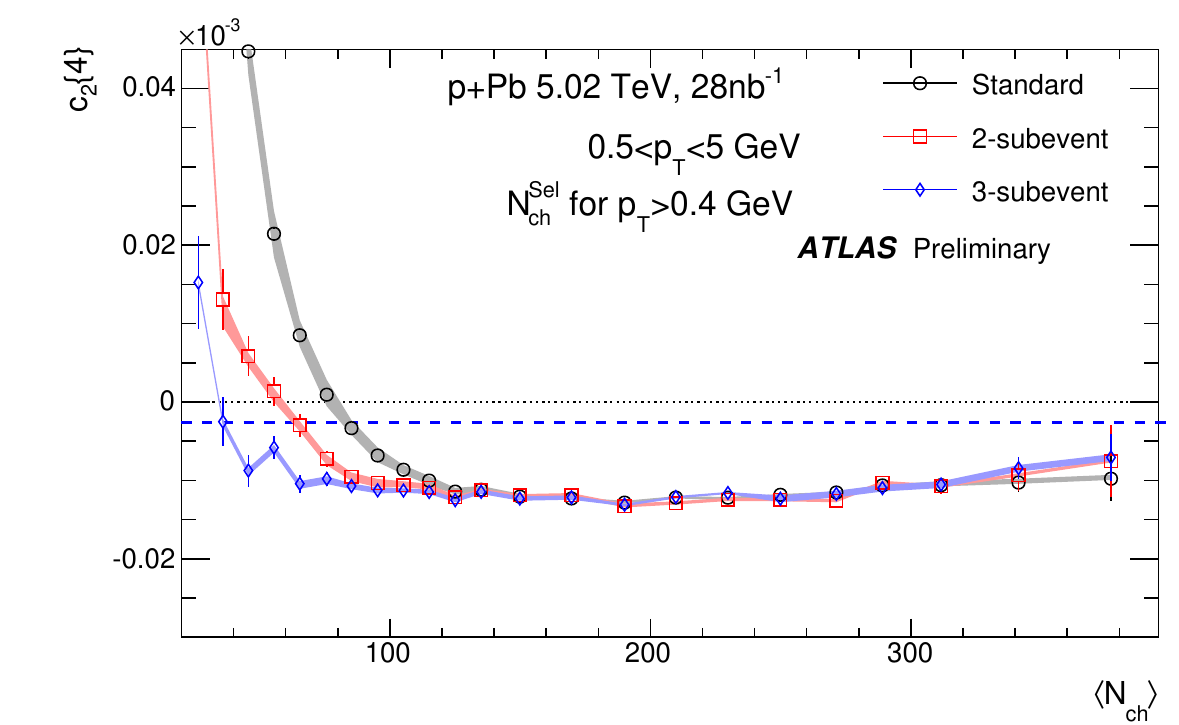}
\caption{The two- and three-subevent methods~\cite{jiangyong} were used to measure $c_2\{4\}(\langle N_{ch}\rangle)$ in (left) \pp{} and (right) \pPb{} collisions at $\sqrt{s}$ = 13 TeV and $\sqrts$ = 5.02 TeV, respectively, in ATLAS~\cite{ATLASv24subevents}.\label{fig:v24subevents}}
\end{figure}

The three-subevent method was used to measure $v_2\{4\}$ in \pPb{} and \pp{} collisions, and it was shown to be more robust with respect to several choices of multiplicity estimator than the standard method.  As shown in Fig.~\ref{fig:v24subevents}, in \pPb{} the same results are obtained with the standard method as with the two- and three-subevent methods, except at low multiplicities where non-flow is suppressed by the subevent technique.  In \pp{} collisions, non-flow is significant at all multiplicities but the three-subevent method yields significant negative $c_2\{4\}$ in all but the lowest-multiplicity collisions.  Four- and six-particle cumulants were also used to obtain $v_2\{4\}$ and $v_2\{6\}$ in \pp{} collisions in~\cite{CMSppcollectivity}, in which also a mass ordering of the $v_2$ for $K^0_S$ mesons and $\Lambda$ baryons was observed using the subtraction method.  While techniques for reducing non-flow are still rapidly developing, these new methods for measuring $c_n\{4\}$ represent a significant advancement in our investigation of collective effects in \pp{} collisions.  

\section{Conclusions \& Outlook}

The ridge and resulting $v_n$ coefficients have been observed and measured in many collision systems, from proton-proton collisions to asymmetric collisions (including \pAl{}, \pAu{}, \pPb{}, \dAu{}, and \HeAu{}) to heavy-ion collisions (\AuAu{}, \PbPb{}), and across a wide range of center-of-mass energies, from $\sqrts$ = 0.02 to 13 TeV.  Many of the features observed in the small collision systems are reminiscent of those in heavy-ion collisions, such as the existence of a double ridge leading to measurable non-zero $v_n$ coefficients for both light- and heavy-flavor hadrons and the mass ordering of the $v_n$ coefficients.  Multi-particle cumulant measurements confirm that the $v_n$ coefficients reflect true multi-particle correlations, instead of a small number of highly-correlated particles (such as those resulting from jet fragmentation), although the experimental removal of non-flow effects remains very challenging particularly in \pp{} collisions.  

There are also new developments in the study of small systems on the horizon: for example, in heavy-ion collisions the symmetric cumulants (SC) are proposed to be more sensitive to the initial state and transport properties of the medium than the $v_n$ coefficients alone~\cite{ALICEsc}.  The first measurements of SC(2,3) and SC(2,4) have been performed in \pPb{} and \pp{} collisions~\cite{CMSmaxime}, and while the contribution of non-flow still needs to be carefully assessed, these observables may also give more information about the dynamics of collective behavior in small systems as well.  

The presence of collective effects in small systems -- where collectivity is defined as multi-particle correlations across pseudorapidity due to a common source -- is becoming more established thanks to the wealth of experimental results produced in recent years.  However, there are still important questions regarding the interpretation of these observations: Is the mechanism for producing $v_n$ coefficients in small systems the same as in heavy-ion collisions?  Which theoretical models can explain the full set of experimental data?  What does the presence of collectivity in small systems mean for our interpretation of heavy ion collisions?  Just what is behaving collectively in small systems (keeping in mind that the explanation must be consistent with measurements from other sectors such as jet and heavy flavor physics)?  These questions and others will continue to be addressed by new theoretical and experimental work in the near future.  

\section*{Acknowledgements}
This work has been supported by BMBF and SFB 1225 ISOQUANT.

\bibliographystyle{elsarticle-num}
\bibliography{aohlson_qm2017proc}


\end{document}